\documentclass[%
 reprint,showkeys,
superscriptaddress,
 amsmath,amssymb,
 aps,prl
]{revtex4-2}
\usepackage{subcaption} 
\usepackage{graphicx}% Include figure files
\usepackage{dcolumn}% Align table columns on decimal point
\usepackage{bm}% bold math
\begin{document}

\preprint{APS/123-QED}

\title[Feeding a Kerr black hole with quantized vortices]{Feeding a Kerr black hole with quantized vortices}% Force line breaks with \\

\author{Shilong Jin}
\affiliation{MCCTP, Minjiang University, Fuzhou, 350108, China}
\affiliation{Center for Applied Mathematics and KL-AAGDM, Tianjin University, Tianjin, 300072, China}

%\email{jinshilong2000@tju.edu.cn}
\author{Xiaofei Zhao}
%\email{matzhxf@whu.edu.cn}
\affiliation{School of Mathematics and Statistics \& Hubei Key Laboratory of Computational Science, Wuhan University, Wuhan, 430072,  China}

\author{Yong Zhang}
%\email{Zhang\_Yong@tju.edu.cn}
\affiliation{Center for Applied Mathematics and KL-AAGDM, Tianjin University, Tianjin, 300072, China}

\author{Chi Xiong}
%\email{xiongchi@mju.edu.cn}
\affiliation{MCCTP, Minjiang University, Fuzhou, 350108, China}


%\date{\today}% It is always \today, today, %  but any date may be explicitly specified

\begin{abstract}
By solving a nonlinear Klein-Gordon equation in Kerr geometry, we uncover new phenomena and key characteristics of quantized vortices in quantum fluids near a Kerr black hole. The formation of these vortices induces rotational or turbulent flows, which profoundly alter the fluid properties and revise those dark matter models describing axion condensates, ultralight boson clouds, and other scalar fields in the vicinity of spinning black holes. As macroscopic, quantum, and topological defects, these vortices can stably orbit the black hole over extended periods, establishing their viability as novel probes for investigating black hole physics. For instance, we calculate the angular velocities of orbiting vortices to quantitatively characterize the frame-dragging effect, a classic prediction of general relativity. Additionally, we observe that relatively large vortices are accreted onto the black hole, wrapping around it while undergoing splitting and reconnecting processes. In quantum fluids with high vortex densities, turbulent flows emerge, accompanied by the formation of a vortex boundary layer near the event horizon. Beyond the ergosphere, we find vortex emissions and energetic outbursts, which may provide crucial insights into analogous astrophysical events recently discovered by the XRISM satellite.

%\begin{description}
%\end{description}
\end{abstract}

\pacs{Valid PACS appear here}% PACS, the Physics and Astronomy % Classification Scheme.
%\keywords{Suggested keywords}%Use showkeys class option if keyword display desired

                              
\maketitle

%\tableofcontents
%\section{Section I}

Quantum fluids -- including superfluids and Bose-Einstein condensates (BECs) -- are not only the subject of intensive laboratory investigation, but are also hypothesized to exist in celestial objects and galactic haloes. For instance, neutron stars are widely believed to harbor a superfluid core composed of neutron pairs \cite{Chamel-17}; BEC-type cold dark matter (BEC-CDM) is postulated to consist of very light bosons such as QCD axions, axion-like particles and ultra-light bosons, with masses ranging from about $1$ eV down to $10^{-24}$eV \cite{Sikivie-83, Press-90, Sin-94, Lee-96, Hu-00, Matos-01, Amendola-06, Bohmer-07, Sikivie-09, Shapiro-12, Huang-14, Schive-14, Xiong-14, Good-16, Witten-17, Berezhiani-23, Ferreira-21, Matos-24}. Recent advances in this field motivate us to explore the behavior of such quantum fluids in the vicinity of spinning black holes. It has been proposed that boson clouds can form around spinning black holes through the superradiant instability \cite{Brito-15}, and their growth and dissipation processes have been leveraged to probe the ultralight bosons via gravitational wave or electromagnetic signature measurement, such as the searches conducted by the LIGO and Virgo collaborations \cite{Arvanitaki-17, Ng-21, LIGO-22}. 

However,  previous studies on BEC-CDM and boson clouds around spinning black holes have largely neglected quantized vorticity -- a key characteristic of quantum fluids \cite{Onsager-49, Feynman-55, Donnelly-91} that is directly tied to the angular momentum of spinning black holes. Quantized vortices govern the intrinsic properties of quantum fluids, analogous to how their gauged counterparts, Abrikosov vortices define Type-II superconductors. While quantum vortices can be experimentally produced in rotating superfluid helium or BEC of cold atoms, two critical questions arise: can vortices emerge in the quantum fluids when spacetime itself is rotating, and how do they behave under extreme gravity and rapid rotation? A spinning Kerr black hole provides an ideal geometric backdrop to address these questions, where the role of quantized vortices is a macroscopic, quantum and topological defect interacting with the black hole. Notably, similarities between quantized vortices and magnetic fluxes imply that the former exert a substantial influence on dark matter condensates or boson clouds, much like the latter modulate black hole accretion disks and jets. Furthermore, vortices can form diverse structures -- including vortex lattices, boundary layers, vortex tangles, and turbulence -- potentially driving a paradigm shift in our comprehension of BEC-CDM  and boson cloud models. Here, we report new phenomena and key characteristics of quantized vortices near a Kerr black hole, establishing their viability as a novel black-hole physics probe with inherent quantum numbers.

%\section{NLKG with Kerr geometry}

\begin{figure}[h]
	\centering
	\begin{subfigure}{0.2\textwidth}
		\captionsetup{skip=2pt}
		\centering
		\includegraphics[width=\linewidth]{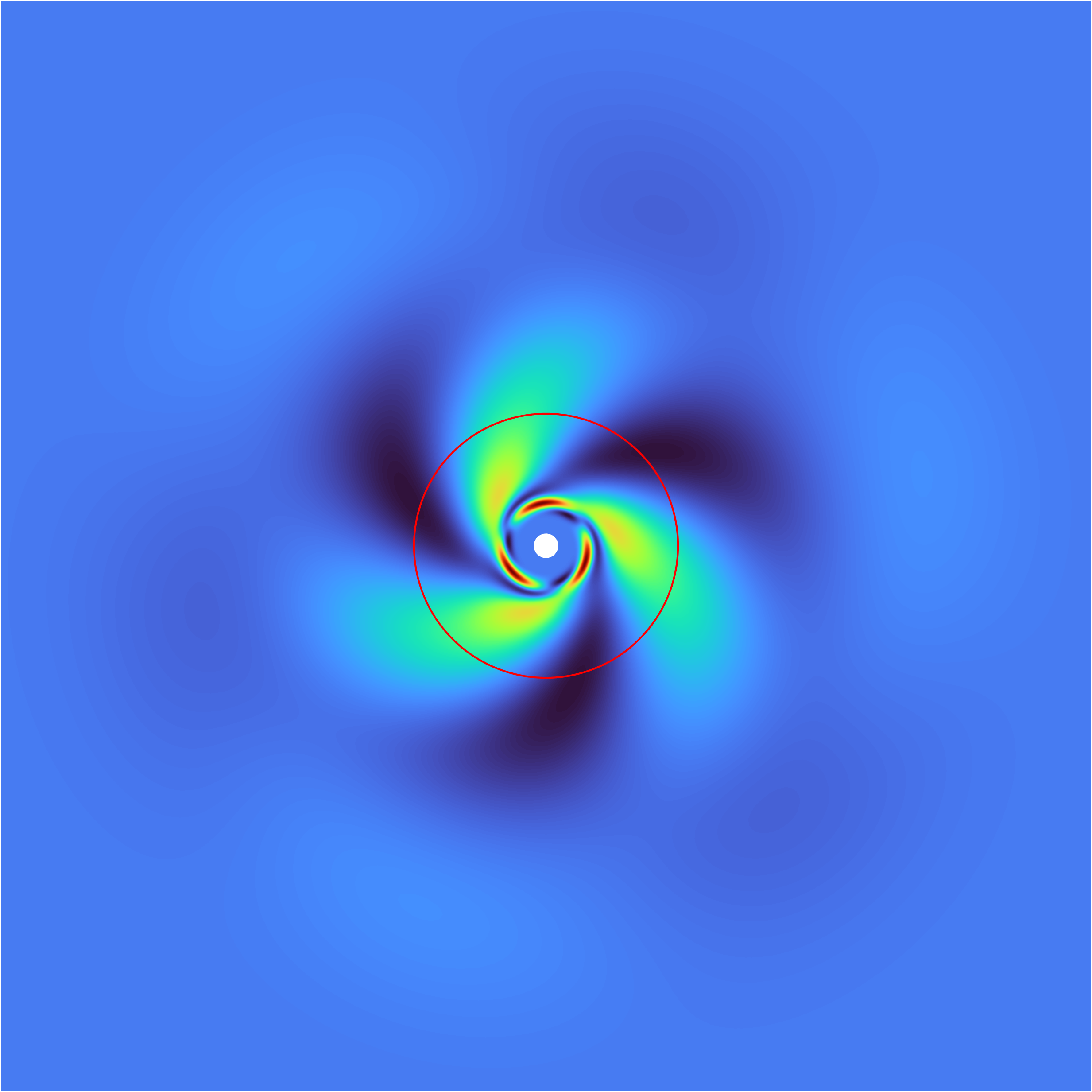}
		\caption{} 
	\end{subfigure}
	\hspace{0.0001\linewidth}
	\begin{subfigure}{0.2\textwidth}
		\captionsetup{skip=2pt}
		\centering
		\includegraphics[width=\linewidth]{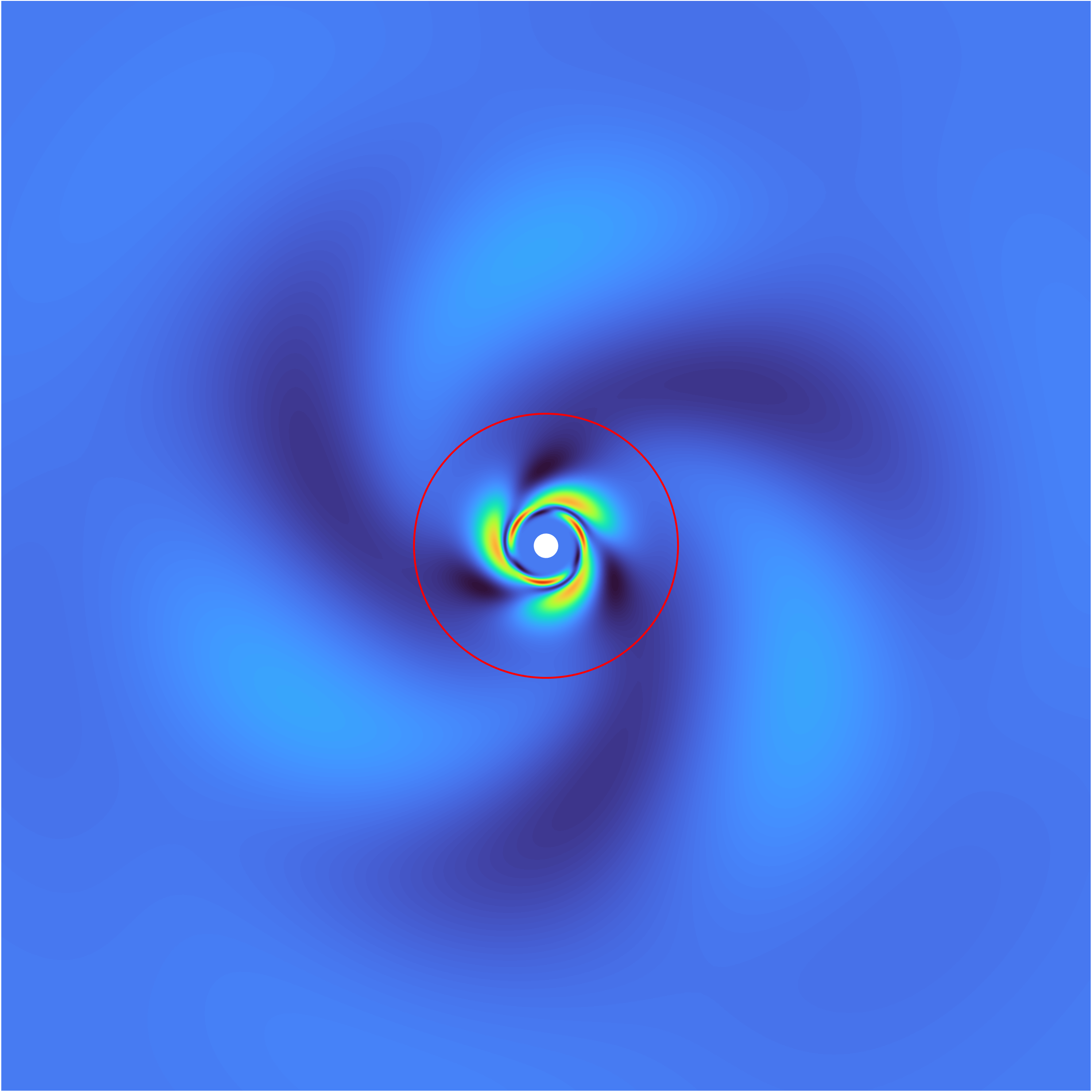}
		\caption{} 
	\end{subfigure}

	\begin{subfigure}{0.2\textwidth}
		\captionsetup{skip=2pt}
		\centering
		\includegraphics[width=\linewidth]{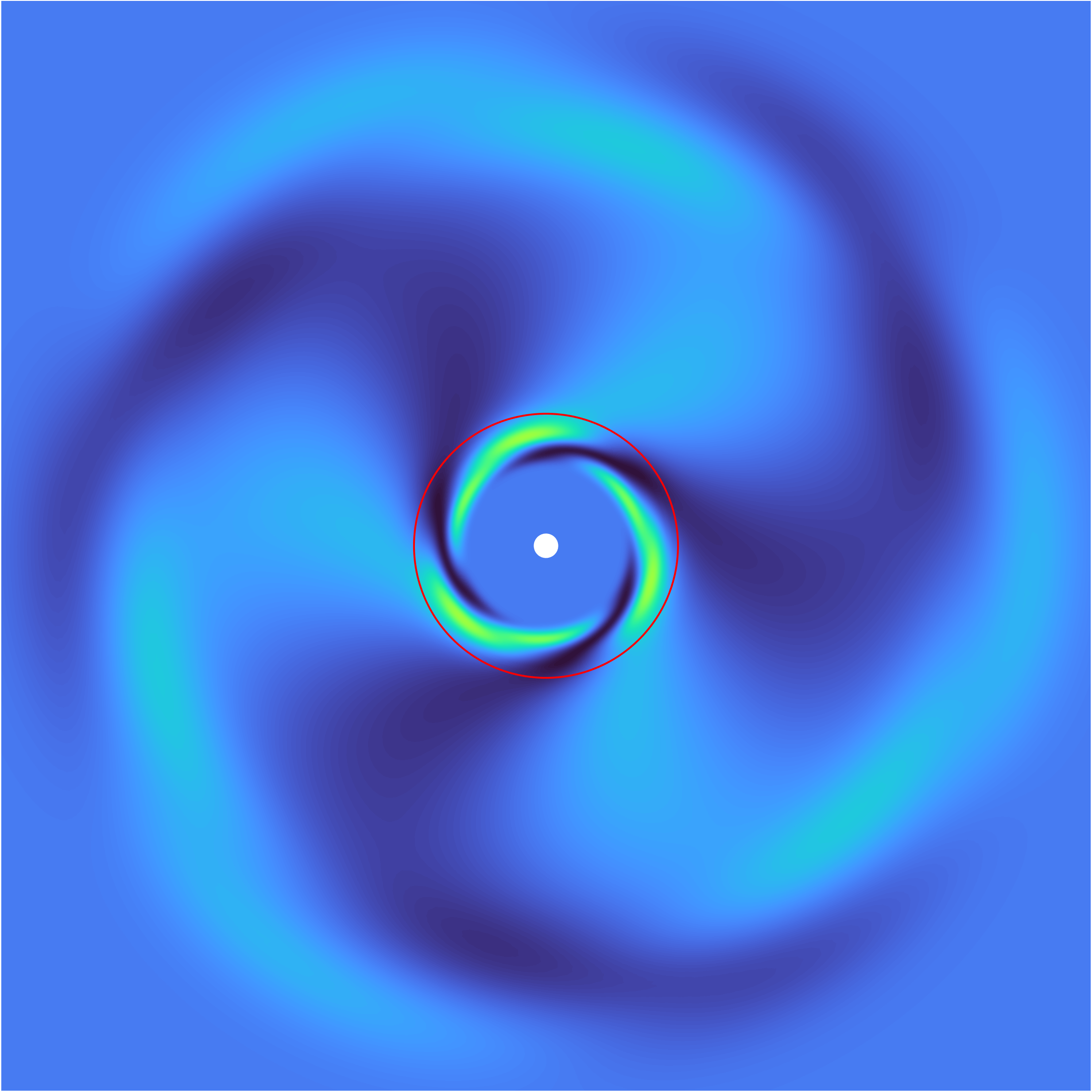}
		\caption{} 
	\end{subfigure}
	\hspace{0.0001\linewidth}
	\begin{subfigure}{0.2\textwidth}
		\captionsetup{skip=2pt}
		\centering
		\includegraphics[width=\linewidth]{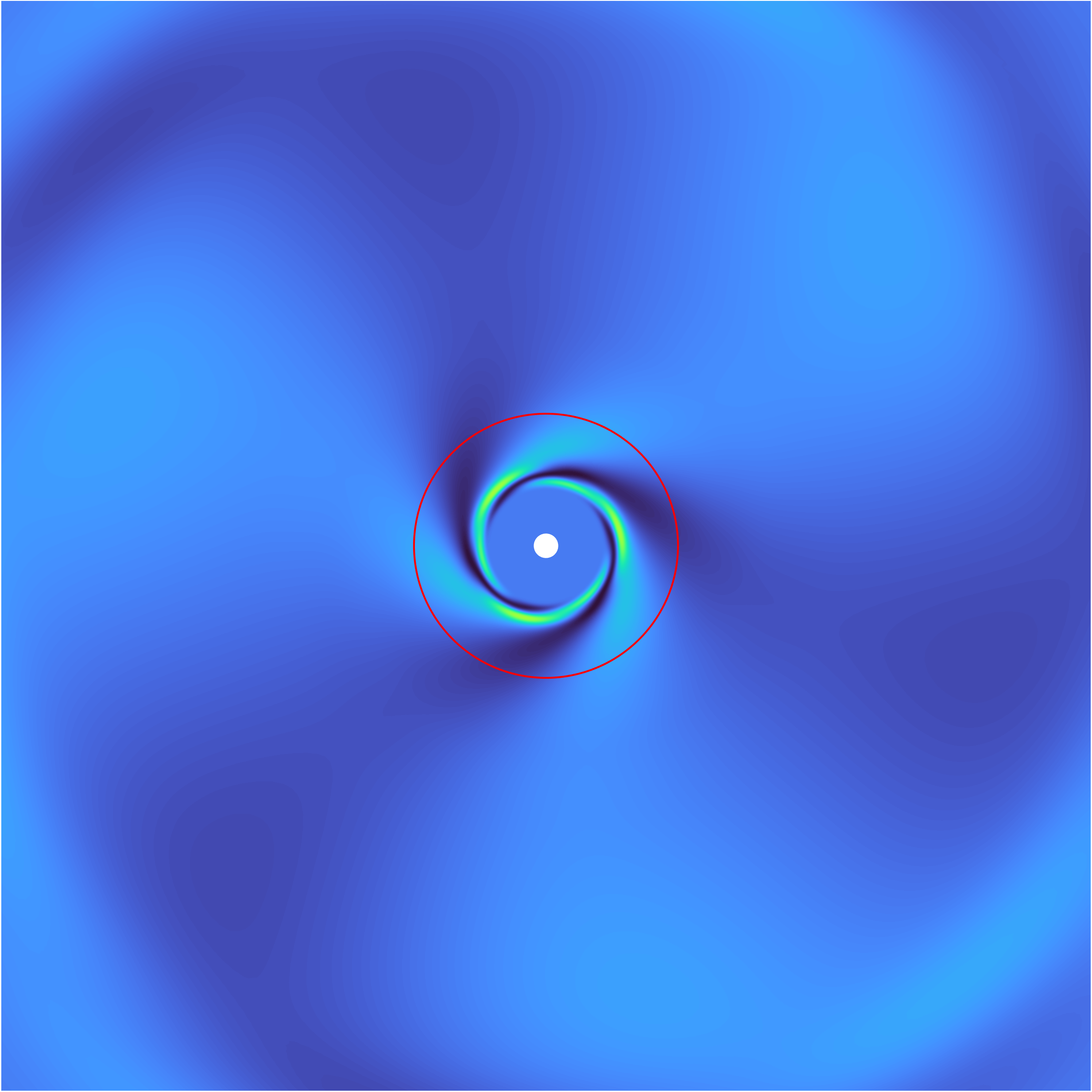}
		\caption{} 
	\end{subfigure}
	\vspace{-0.1cm} 
	\caption{\small{Density plots of $|\Phi|^2$ at the equatorial plane. $\delta \equiv 1- a = 10^{-7},  \lambda=0.1$. The static limit of the ergosphere is indicated by the red circles. (a) and (b): direct rotation -- the fluid follows the rotation of the black hole (counter-clockwise) when spiralling to the event horizon;  (c) and (d): retrograde rotation -- the fluid rotates clockwise initially, and then splits into two parts rotating in opposite directions when entering the ergosphere.}}\label{fig1}
\vspace{-0.5cm}	
\end{figure}

{\it NLKG in Kerr geometry} -- We introduce a complex scalar field, $\Phi = |\Phi| e^{i\sigma}$, to serve as the order parameter of a relativistic quantum fluid in the vicinity of a Kerr black hole. The evolution of the system is governed by a nonlinear Klein-Gordon equation (NLKG) in the Kerr background, $\Box_{\textrm{\tiny{Kerr}}} \Phi = \mathrm{d}V/\mathrm{d}\Phi^*$, where the d'Alembertian is built on the Kerr metric and the nonlinear potential $V(\Phi, \Phi^*) = \lambda/2\,(|\Phi|^2 - F^2_0)^2$, where $F_0$ is the vacuum expectation value of $\Phi$. A quantized vortex is a solution to the NLKG with a constraint $\oint_C \partial_\mu  \sigma \mathrm{d}x^\mu = 2 \pi n$ over some spatially closed loop $C$, where $n$ is the winding number. Noticing that it only involves the derivative of the phase of $\Phi$, we use this constraint directly to identify vortices.  

One can first take the slow-rotation limit to obtain an intelligible picture and a connection to the case of a Schwarzschild black hole.  At this limit ($a \equiv J/M \ll 1$, where $J, M$ are the angular momentum and the mass of the Kerr black hole, respectively), the ergosphere of the black hole, which is the region between the event horizon ($r=r_{+}$) and the static limit ($r=r_{+}^{E}$), becomes a thin rotating shell with coordinate angular velocity $\Omega \approx a/r_{+}^{2}$. This provides a gravitational analogue of the rotating laser traps for confining the BEC of cold atoms -- the rotating ``bucket" here is the spacetime itself. With a small-$a$ expansion, one can show that $\Box_{\textrm{\tiny{Kerr}}} \Phi \approx \Box_{\textrm{\tiny{Schw}}} \Phi - 2\Omega/g_{tt} \partial_{\phi t} \Phi +\Omega^2/g_{tt} \partial_{\phi \phi} \Phi$, where the d'Alembertian $\Box_{\textrm{\tiny{Schw}}}$ is built on the Schwarzschild metric, and the last two terms are identified as the ``Coriolis" term and the ``centrifugal" term, respectively. These two terms represent the ``inertial forces" due to the black hole rotation and also illustrate Mach's principle. Consequently, quantized vortices and vortex lattices can be produced by the Coriolis term, which reduces to the usual rotating term $\sim \Omega L_z \psi$ as the NLKG reduces to the nonlinear Schr\"{o}dinger equation for the order parameter $\psi$ at the nonrelativistic limit \cite{Xiong-14, Zhao-20}. For fast-spinning black holes, the small-$a$ expansion fails. Therefore we solve the NLKG numerically to find exact solutions in the following investigations. 

Quantized vortices deplete the density in their core region and hence, significantly modify the density profile of a quantum fluid. We start with a numerical solution without quantized vortices: under appropriate initial conditions, a quantum fluid evolves into three high-density components that rotate around the black hole with extended arms stretching out as they spiral toward the event horizon. Fig.\ref{fig1} shows the density profiles of the fluid for two tests, which are conducted by releasing the high-density components of the fluid near the ergosphere in direct and retrograde rotations, respectively. Motions due to the black hole's gravitational pull and rotation are apparent in both cases, and manifest the frame-dragging effect for Kerr black holes. 

\begin{figure*}
	\centering
	
		\begin{subfigure}{0.28\textwidth}
		%		\captionsetup{skip=0pt}
		\centering
		\includegraphics[width=0.87\linewidth]{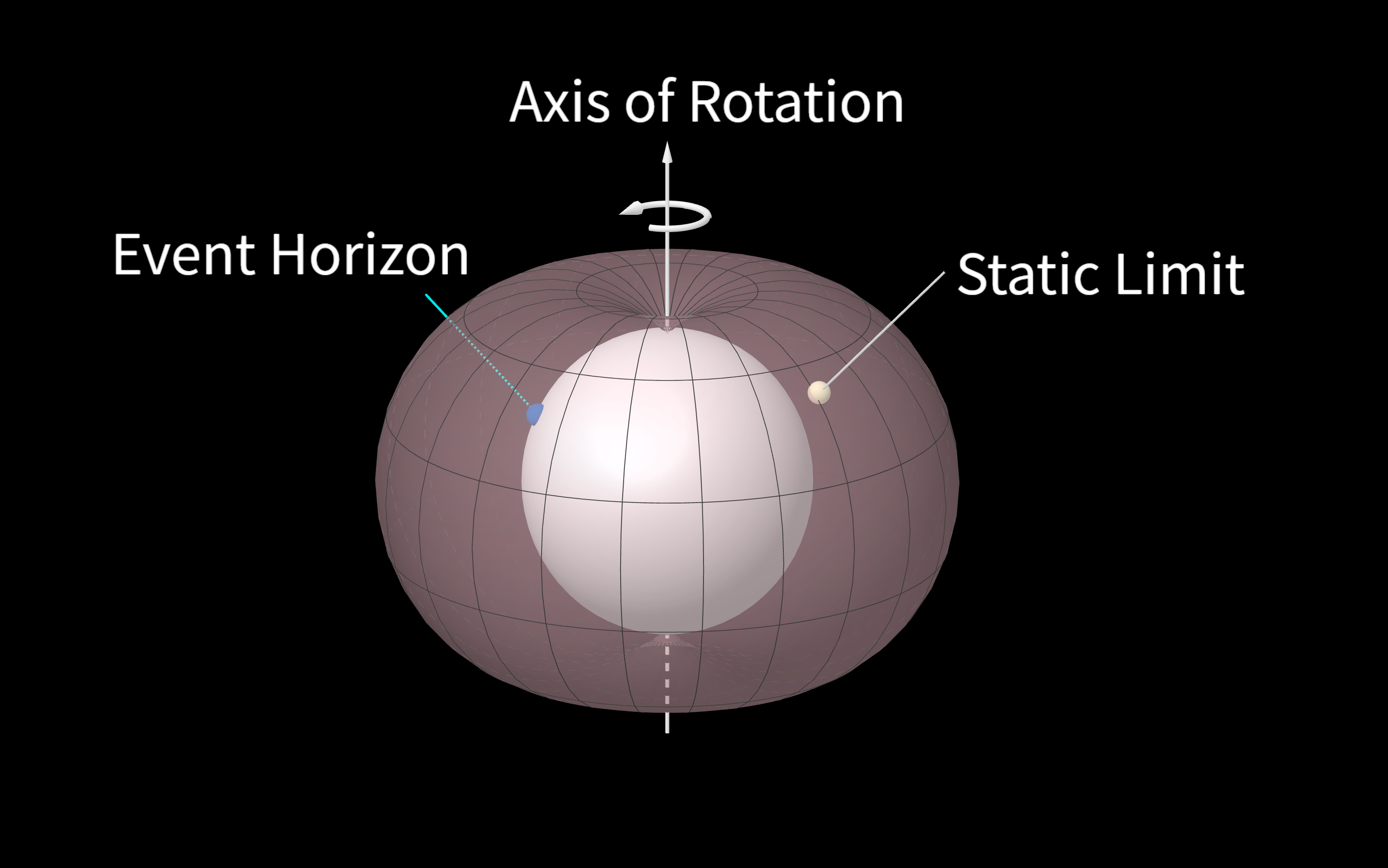}
		\caption{} \label{Kerr}
	\end{subfigure}
	\hfill	
	\begin{subfigure}{0.38\textwidth}
		%		\captionsetup{skip=-10pt}
		\centering
		\includegraphics[height=2.85cm]{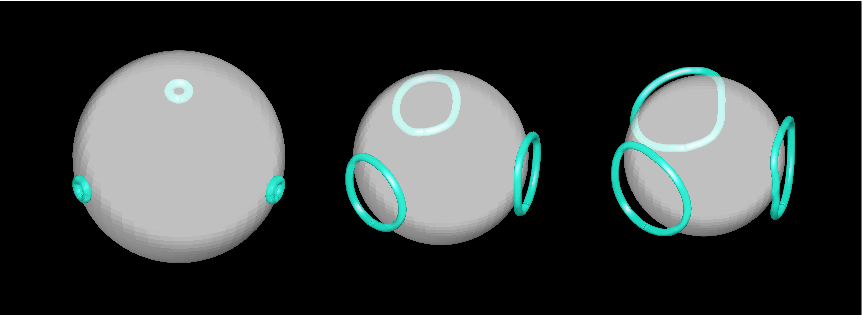}
		\caption{} \label{fig2_diff_a}
	\end{subfigure}
	\hfill
	\begin{subfigure}{0.2\textwidth}
		%		\captionsetup{skip=0pt}
		\centering
		\includegraphics[height=2.85cm]{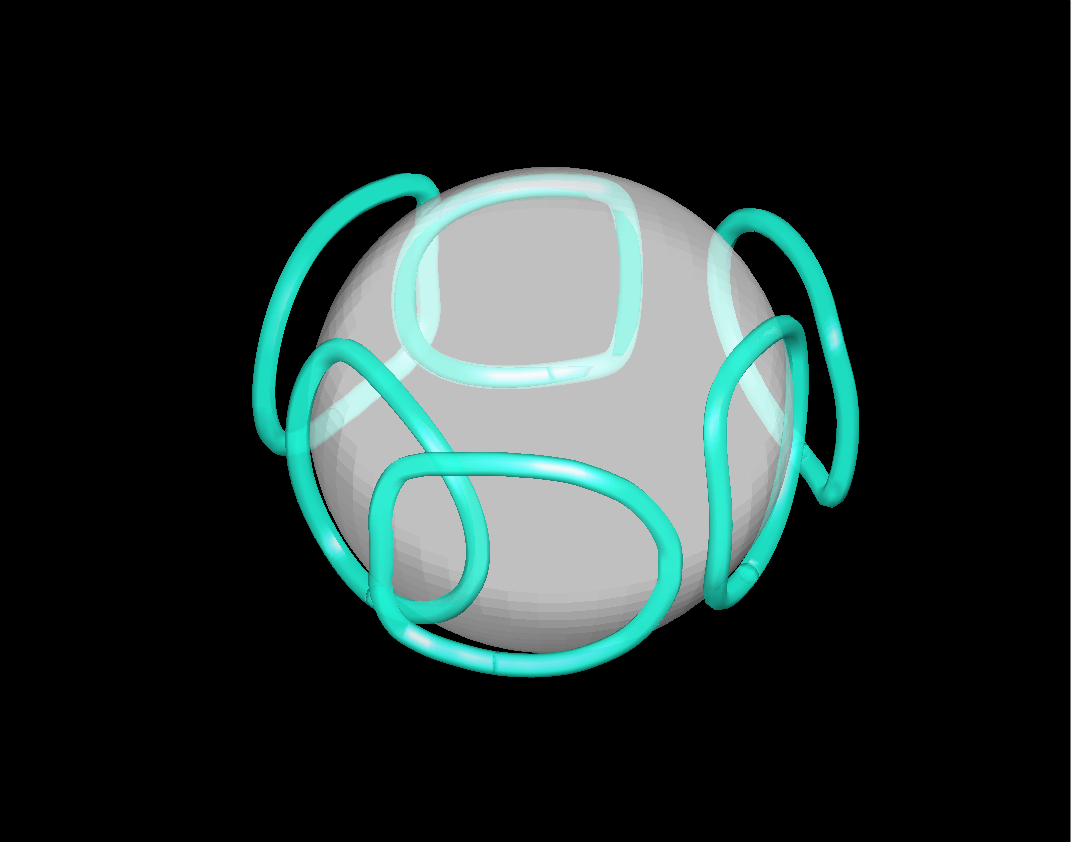}
		\caption{} \label{fig2_2layer}
	\end{subfigure}

	\begin{subfigure}{0.36\textwidth}
		%		\captionsetup{skip=0pt}
		\centering
		\includegraphics[height=3.0 cm]{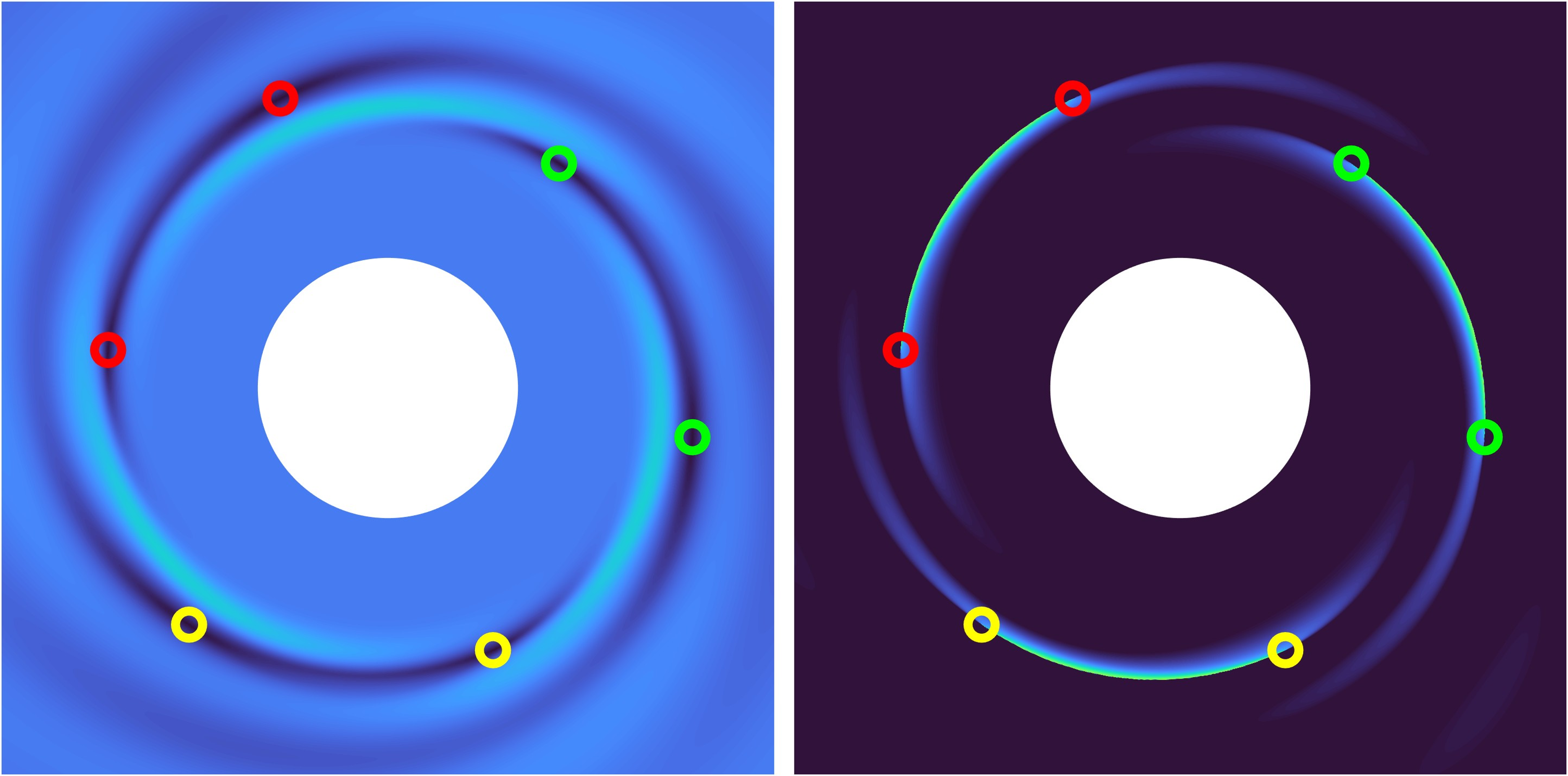}
		\caption{} \label{fig2_v3_dense}
	\end{subfigure}
	\hfill 
	\begin{subfigure}{0.62\textwidth}
		%		\captionsetup{skip=-10pt}
		\centering
		\includegraphics[height=3.0 cm]{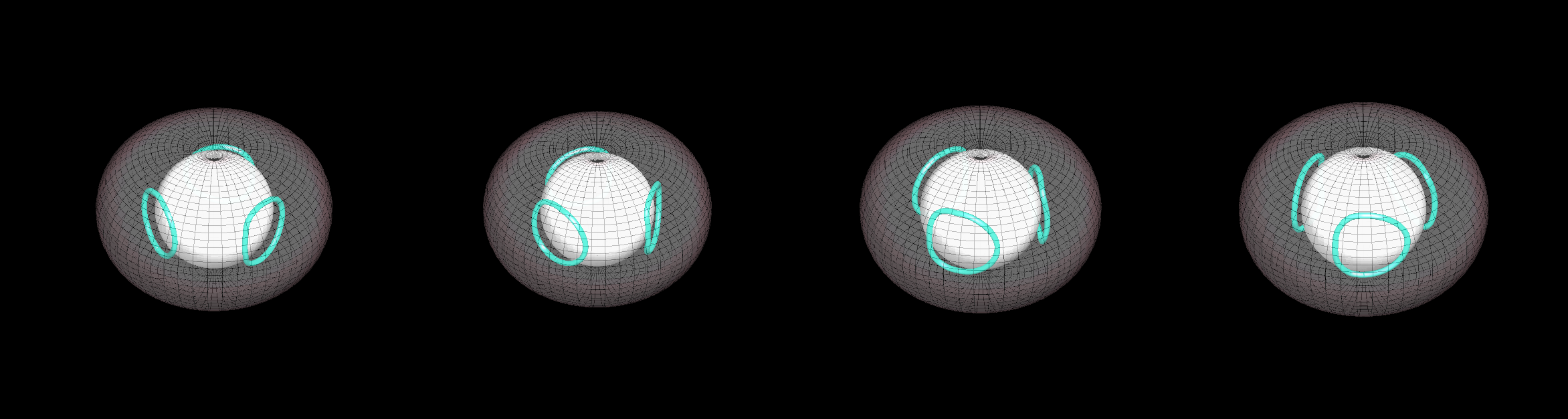}
		\caption{} \label{fig2_v3_T}		
	\end{subfigure}

	\caption{\small{Quantized vortices orbiting around a Kerr black hole. The event horizons are indicated by the grey surfaces in (a-c, e) and the white circles in (d), respectively. (a) An illustration for a Kerr black hole: the ergoshpere is the region enclosed by the grey and brown surfaces;  (b) Three cases with increasing rotational velocities of the black hole which correspond to $\delta = 10^{-2}, 10^{-3}$ and $10^{-7}$, respectively ($\lambda=0.1$); (c) Six vortices around the black hole; (d) Density (left) and phase (right) plots of $\Phi$ on the equatorial plane: Positions of each vortex intersecting with the equatorial plane are labelled by circles of the same color; (e) Three vortices orbit the black hole. }}\label{fig2}
\vspace{-0.5cm}	
\end{figure*}

\begin{figure}[h!]
	\centering
	\includegraphics[width=0.450\textwidth]{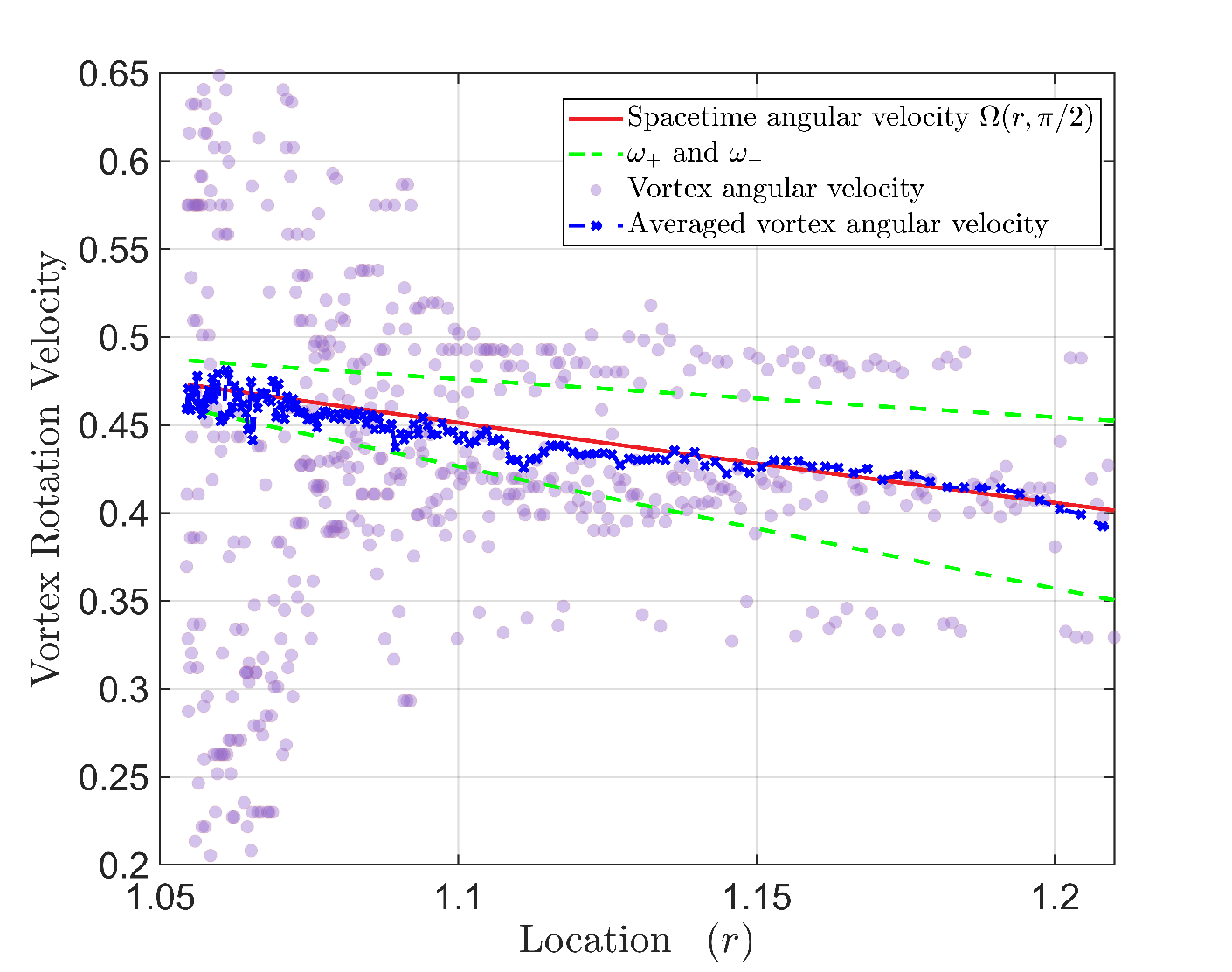}
	\caption{\small{The averaged angular velocities of three orbiting vortices (blue crosses) matches with the angular velocities of locally non-rotating observers (red line). }
	}\label{fig3}
\vspace{-0.5cm}	
\end{figure}

%\section{Vortex orbiting}
{\it Vortex orbiting} -- Besides the density $|\Phi (x)|$, information about the phase of the order parameter, $\sigma(x)$, also requires presentation and visualization, like the trajectory and velocity of a test particle ``fed" to the black hole. Nevertheless, the wave function nature of $\Phi$ hinders identifying an {\it object} that fulfills the test particle's role. This changes with the appearance of quantized vortices: their core centers correspond to density zeros and phase singularities, around which $\sigma(x)$ varies from 0 to $2\pi n$ along any closed contour threaded by the vortices; they are not only topological defects with winding numbers $n$, but also macroscopic entities encoding both the density and phase information of $\Phi$.  Figs.\ref{fig2_v3_dense} and \ref{fig2_v3_T} visualize a solution with three quantized vortex rings orbiting within the black hole's ergosphere and gradually approaching the event horizon. This stable orbital behavior -- persisting for extended period -- resembles the vortex lattice in a rotating BECs or superfluids. Just as a test particle traversing its geodesic probes spacetime properties, the unique characteristics of quantized vortices establishing them as novel detectors for black holes (and potentially other curved spacetimes). Here we give a quantitative example, using the above orbiting vortices to mimic the so-called locally non-rotating observers with coordinate angular velocity $\Omega = \mathrm{d}\phi/\mathrm{d}t = - g_{t\phi}/g_{\phi\phi}$, and other stationary observers outside the event horizon with coordinate angular velocity $\omega$ ($\omega_{-} < \omega < \omega_{+}$), where $\omega_{\pm} = \Omega \pm \sqrt{\Delta} \sin\theta/g_{\phi\phi}$ and $ \Delta = r^2 + a^2 - 2 M r$. Fig. \ref{fig3} summarizes the measurement of $\omega$ values by tracing the orbital motion of a vortex lattice, which not only illustrates the frame-dragging effect, but also verifies the inequality $\omega_{-} < \omega < \omega_{+}$ quantitatively. The stability and precise locatability of quantized vortices render them exceptional probes for investigating black hole physics. 

\begin{figure}[h!]
	\centering
	\includegraphics[width=0.45\textwidth]{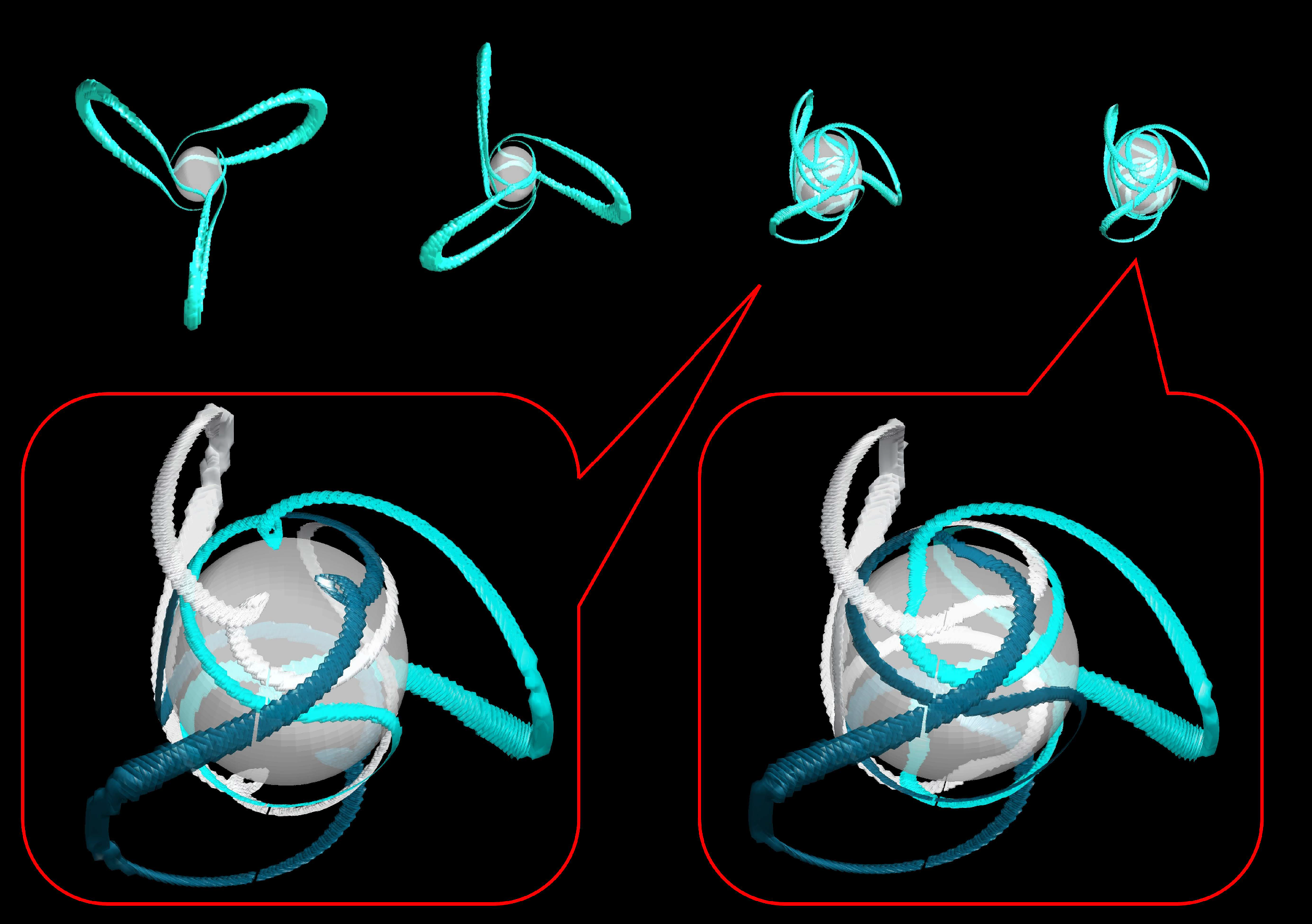}
	\caption{\small{Three relatively large vortices wind on the spinning black hole and reconnect to form a trefoil-like knot.}}\label{fig4}
\vspace{-0.5cm}		
\end{figure}

%\section{Spaghettification and winding}

{\it Spaghettification and winding} -- How do quantized vortices react to the tidal force when they are ``fed" to a black hole? Notably, a quantized vortex can be effectively modeled as a spinning string with tension \cite{Davis-89}, which should stretch under the black hole's strong gravitational pull, similar to the ``spaghettification" of normal objects  (e.g., stars) venturing too close. Could tidal forces tear it apart? Such a scenario would contradict the fact that a quantized vortex must terminate on a boundary or form a closed ring \cite{Onsager-49, Feynman-55, Donnelly-91}; otherwise they collapse completely and the superfluidity or condensation is destroyed. As illustrated earlier, small vortex rings exhibit long-term stability against tidal forces, though shape deformation and length variation are not visually evident. To gain more insight into this question, three relatively large vortices are released from regions outside the ergosphere.  Fig.\ref{fig4} depicts how these quantized vortices ``wind" around the spinning black hole, analogous to thread on a yarn ball winder.  Intriguingly, these vortices become tangled as they co-rotate with the black hole, and reconnect in the polar area above the event horizon, forming a trefoil knot-like structure (Fig.\ref{fig4}). Our observations might relate to studies on cosmic strings interacting with spinning black holes \cite{Vilenkin-23}, and local strings produced by black hole superradiance \cite{East-22}. While quantized vortices and cosmic strings share both similarities and differences in their interactions with black holes \cite{Davis-89}, a detailed comparison will be presented in a separated work \cite{Jin-25}.  

%\section{Vortex tangle, boundary layer and emissions}

\begin{figure*}
	\centering
	\begin{subfigure}{0.6\textwidth}
		%		\captionsetup{skip=-10pt}
		\centering
		\includegraphics[width=\linewidth]{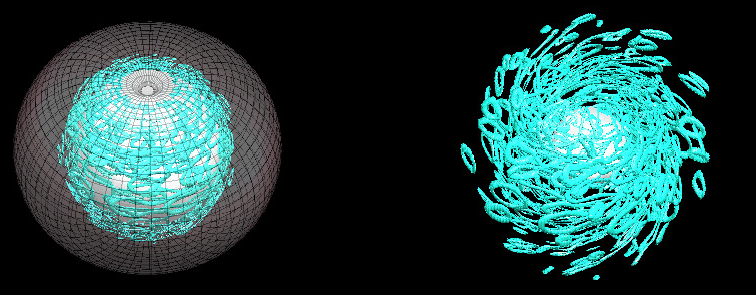}
		\caption{} \label{fig5_1}
	\end{subfigure}
	
	\begin{subfigure}{1.0\textwidth}
		%		\captionsetup{skip=0pt}
		\centering
		\includegraphics[width=\linewidth]{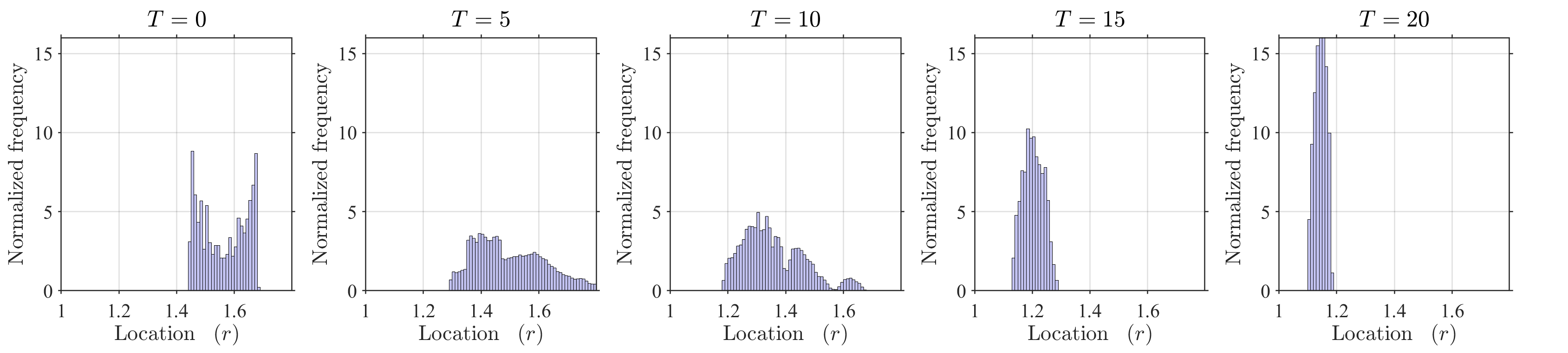}
		\caption{} \label{fig5_2}
	\end{subfigure}
	
	\caption{\small{Turbulent states in the ergosphere: (a) A vortex ``lantern" with openings at the polar region of the black hole. The picture on the right, which visualizes more details of the turbulent flow, is connected to the left one by a coordinate transformation; (b) A histogram plot of the vorticity distribution inside the ergosphere: Initially located at the static limit, vortices advect toward the event horizon, forming a narrower and denser layer there.}}\label{fig5}
	\vspace{-0.2cm}	
\end{figure*}

\begin{figure*}
	\centering
	\begin{subfigure}{0.25\textwidth}
		%		\captionsetup{skip=-10pt}
		\centering
		\includegraphics[height=3.1cm]{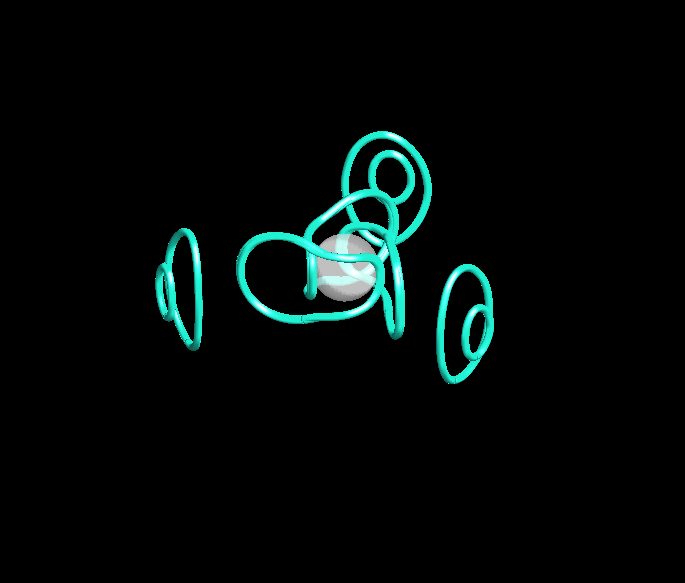}
		\caption{} 
	\end{subfigure}
	\hfill
	\begin{subfigure}{0.73\textwidth}
		%		\captionsetup{skip=0pt}
		\centering
		\includegraphics[height=3.1cm]{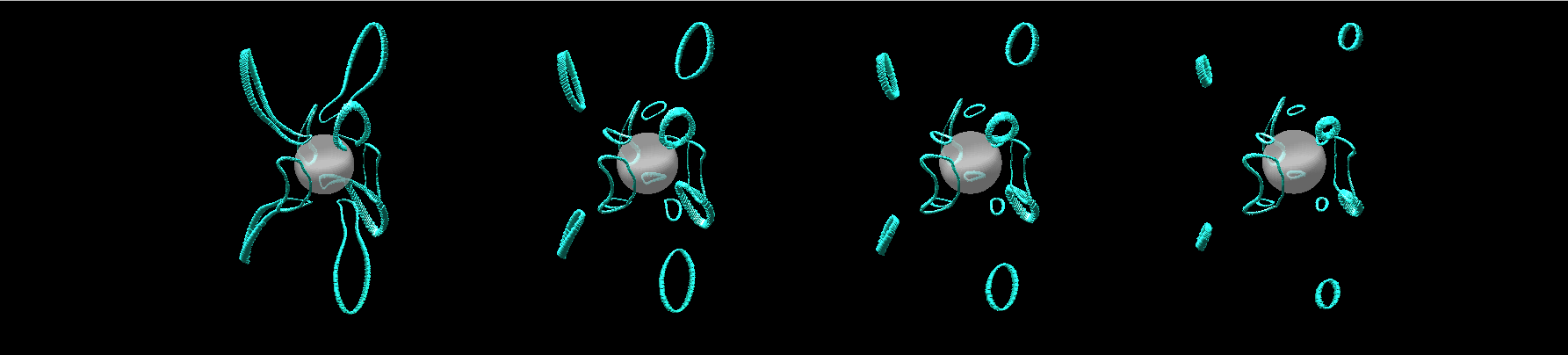}
		\caption{} 
	\end{subfigure}
	
	\caption{\small{(a) Emission of vortices at the equatorial plane. Three relatively smaller vortices are separated from the larger ones and move outwards while shrinking. A subsequent snapshot is superposed to show the relative positions of emitted vortices; (b) Formation and emission of vortex rings at high latitude positions ($\theta \approx 57\,^{\circ}$).}}\label{fig6}
	\vspace{-0.5cm}	
\end{figure*}

{\it Vortex tangle, boundary layer, and emission} -- So far, our computations have focused on systems with a small number of vortices. However, the substantial angular momentum of a spinning black hole may induce the production of an enormous number of vortices. Similar to superfluid helium or BECs -- where vortex splitting and reconnection derive vortex tangles and continuous or intermittent quantum turbulence \cite{Feynman-55, Schwarz-88} -- these processes also occur to in quantum fluids near Kerr black holes. For flat spacetimes, relativistic quantum turbulence have been simulated and visualized by solving the NLKG with the Minkowski metric \cite{Liu-21}. Here, we present numerical simulations of vortex tangles and turbulent flows in quantum fluids around Kerr black holes. Fig.\ref{fig5} demonstrates that producing numerous vortices in the ergosphere leads to the formation of  a ``lantern-shaped" vortex tangle above the event horizon -- effectively a vortex boundary layer. This is further supported by Fig.\ref{fig5_2} which plots the time evolution of vorticity distribution in the ergosphere: quantized vortices are produced near the static limit,  then advect toward the event horizon, forming a vortex boundary layer there. Fig.\ref{fig5} illustrates that the system exhibits an evolutionary trend toward a fully-fledged state of quantum turbulence.  

Lastly, we analyze vortex emission from the ergosphere of a Kerr black hole, analogous to the coronal mass ejections (CMEs) -- explosive bursts of plasma and magnetic field from the Sun's corona.  The formation of the ``lantern-shaped" vortex tangle indicates that with the accumulation of vortices in the ergosphere, the quantum fluid turns into an inhomogeneous turbulent state: vorticity concentrates near the event horizon ($r=r_{+}$) to form a boundary layer, while vortices are emitted outward from the outer ergosphere ($r \sim r_{+}^E$) (Fig.\ref{fig6}), accompanied by energetic outbursts into the black hole's outer space. This scenario resembles both solar CMEs and episodic black hole ejections associated with closed magnetic fluxes in accretion flows \cite{Yuan-09}, and may shed light on recent discovery of XRISM \cite{XRISM-25}: the satellite detected high-speed black hole ``winds" that are clumpy (bullet-like) and omnidirectional -- challenging long-standing galaxy-black hole coevolution theories. We speculate that this clumpy wind structure, with distinct gas components moving at varying velocities, could be linked to vortex emission: vortices may carry gravitationally trapped ions, leaving observable signatures. As shown in Fig.\ref{fig6}, vortex emissions occur at both equatorial and high-latitude regions, consistent with the omnidirectional gas outflows observed by XRISM.

{\it Acknowledgments} -- X. Zhao is supported by the National Key R\&D Program of China No. 2024YFE03240400, National MCF Energy R\&D Program and NSFC 42450275, 12271413. S. Jin and Y. Zhang are partially supported by the National Key R\&D Program of China No. 2024YFA1012803 and basic research fund of Tianjin University under grant 2025XJ21-0010,  National NSFC 12271400. S. Jin and C. Xiong are partially supported by the Startup Grant No. 30804317 from Minjiang University.  

%\section{Conclusions}
%For quantum fluids near a Kerr black hole, we have demonstrated the importance of quantized vortices both on their own merits and as the sources of rotational or turbulent flows. Under the strong gravity and fast rotation of the black hole, the topological stability and macroscopic quantum character of vortices enable them to orbit around the black hole like a chain of satellites, to form a boundary layer over the event horizon when the number of vortices is large,  and to be emitted from the ergosphere with energy outbursts. These features change essentially the basic properties of the fluids near the black hole, similar to how their gauged counterparts, the Abrikosov vortices define the Type-II superconductors. We have explored only a small portion of the parameter space with limited computation scales, and neglected the back reaction to the black hole geometry.  Overcoming these limitation will lead to more remarkable and unanticipated new phenomena including a process for quantized vortices, analogous to the Blandford-Znajek process \cite{Blandford-77} for magnetic fluxes, to extract energy and angular momentum from the black hole, followed by matter and energy emissions to the surrounding galaxy. 

%\nocite{*}

\bibliography{KerrHiggs-bibliography}% Produces the bibliography via BibTeX.

@article{Onsager-49,
    author ={Onsager, L.},
    title = {{Discussion in a paper by C. J. Gorter}},
    journal = {Nuovo Cimento Suppl.},
    volume ={6},
    number ={},
    pages ={249-250},
   year ={1949},
   month ={},
   issn={}
}

@incollection{Feynman-55,
title = {{Chapter II: Application of Quantum Mechanics to Liquid Helium}},
editor = {C.J. Gorter},
series = {Progress in Low Temperature Physics},
publisher = {Elsevier},
volume = {1},
pages = {17-53},
year = {1955},
issn = {0079-6417},
author = {R.P. Feynman}
}

@book{Donnelly-91,
  title={Quantized vortices in helium II},
  author={Donnelly, Russell J},
  volume={2},
  year={1991},
  address   = {Cambridge},
  publisher={Cambridge University Press}
}

@article{Chamel-17,
    author ={{See e.g., Chamel, N.}},
    title = {Superfluidity and Superconductivity in Neutron Stars},
    journal = {Journal of Astrophysics and Astronomy},
    volume ={38},
    number ={3},
    pages ={1-14},
   year ={2017},
   month ={09},
   issn={0973-7758}
}

@article{Sikivie-83,
  title = {Experimental Tests of the ``Invisible" Axion},
  author = {Sikivie, P.},
  journal = {Phys. Rev. Lett.},
  volume = {51},
  issue = {16},
  pages = {1415--1417},
  numpages = {0},
  year = {1983},
  month = {Oct},
  publisher = {American Physical Society},
}

@article{Press-90,
  title = {Single mechanism for generating large-scale structure and providing dark missing matter},
  author = {Press, William H. and Ryden, Barbara S. and Spergel, David N.},
  journal = {Phys. Rev. Lett.},
  volume = {64},
  issue = {10},
  pages = {1084--1087},
  numpages = {0},
  year = {1990},
  month = {Mar},
  publisher = {American Physical Society},
}

@article{Sin-94,
  title = {Late-time phase transition and the galactic halo as a Bose liquid},
  author = {Sin, Sang-Jin},
  journal = {Phys. Rev. D},
  volume = {50},
  issue = {6},
  pages = {3650--3654},
  numpages = {0},
  year = {1994},
  month = {Sep},
  publisher = {American Physical Society},
}

@article{Lee-96,
  title = {Galactic halos as boson stars},
  author = {Lee, Jae-Weon and Koh, In-Gyu},
  journal = {Phys. Rev. D},
  volume = {53},
  issue = {4},
  pages = {2236--2239},
  numpages = {0},
  year = {1996},
  month = {Feb},
  publisher = {American Physical Society},
}

@article{Hu-00,
  title = {Fuzzy cold dark matter: the wave properties of ultralight particles},
  author = {Hu, Wayne and Barkana, Rennan and Gruzinov, Andrei},
  journal = {Phys. Rev. Lett.},
  volume = {85},
  issue = {6},
  pages = {1158--1161},
  numpages = {0},
  year = {2000},
  month = {Aug},
  publisher = {American Physical Society},
}

@article{Matos-01,
  title = {Further analysis of a cosmological model with quintessence and scalar dark matter},
  author = {Matos, Tonatiuh and Arturo Ure\~na-L\'opez, L.},
  journal = {Phys. Rev. D},
  volume = {63},
  issue = {6},
  pages = {063506},
  numpages = {11},
  year = {2001},
  month = {Feb},
  publisher = {American Physical Society},
}

@article{Amendola-06,
title = {Dark matter from an ultra-light pseudo-Goldsone-boson},
journal = {Phys. Lett. B},
volume = {642},
number = {3},
pages = {192-196},
year = {2006},
issn = {0370-2693},
author = {Luca Amendola and Riccardo Barbieri}
}

@article{Bohmer-07,
year = {2007},
month = {jun},
publisher = {},
volume = {},
number = {06},
pages = {025},
author = {B\"{o}hmer, C G and Harko, T},
title = {Can dark matter be a {Bose-Einstein} condensate?},
journal = {Journal of Cosmology and Astroparticle Physics}
}

@article{Sikivie-09,
  title = {{Bose-Einstein} condensation of dark matter axions},
  author = {Sikivie, P. and Yang, Q.},
  journal = {Phys. Rev. Lett.},
  volume = {103},
  issue = {11},
  pages = {111301},
  numpages = {4},
  year = {2009},
  month = {Sep},
  publisher = {American Physical Society},
}

@article{Shapiro-12,
    author = {Rindler-Daller, Tanja and Shapiro, Paul R.},
    title = {Angular momentum and vortex formation in {Bose-Einstein-condensed} cold dark matter haloes},
    journal = {MNRAS},
    volume = {422},
    number = {1},
    pages = {135-161},
    year = {2012},
    month = {04},
    issn = {0035-8711}
}

@article{Huang-14,
author = {Huang, Kerson and Xiong, Chi and Zhao, Xiaofei},
title = {Scalar-field theory of dark matter},
journal = {International Journal of Modern Physics A},
volume = {29},
number = {13},
pages = {1450074},
year = {2014}
}

@article{Schive-14,
Author = {Schive, Hsi-Yu and Chiueh, Tzihong and Broadhurst, Tom},
Title = {Cosmic structure as the quantum interference of a coherent dark wave},
Journal = {Nature Physics},
Year = {2014},
Volume = {10},
Number = {7},
Pages = {496-499},
Month = {JUL},
ISSN = {1745-2473},
EISSN = {1745-2481},
ResearcherID-Numbers = {Broadhurst, Thomas/AAF-9891-2019},
ORCID-Numbers = {Schive, Hsi-Yu/0000-0002-1249-279X
      CHIUEH, TZI-HONG/0000-0003-2654-8763
      Broadhurst, Tom/0000-0002-8785-8979
   },
Unique-ID = {WOS:000338843100015},
}

@article{Xiong-14,
  title = {Relativistic superfluidity and vorticity from the nonlinear {Klein-Gordon} equation},
  author = {Xiong, Chi and Good, Michael R. R. and Guo, Yulong and Liu, Xiaopei and Huang, Kerson},
  journal = {Phys. Rev. D},
  volume = {90},
  issue = {12},
  pages = {125019},
  numpages = {8},
  year = {2014},
  month = {Dec},
  publisher = {American Physical Society},
}

@article{Good-16,
author = {Good, Michael R R and Xiong, Chi and Chua, Alvin J K and Huang, Kerson},
title = {Geometric creation of quantum vorticity},
journal = {New Journal of Physics},
year = {2016},
month = {nov},
publisher = {IOP Publishing},
volume = {18},
number = {11},
pages = {113018}
}

@article{Witten-17,
  title = {Ultralight scalars as cosmological dark matter},
  author = {Hui, Lam and Ostriker, Jeremiah P. and Tremaine, Scott and Witten, Edward},
  journal = {Phys. Rev. D},
  volume = {95},
  issue = {4},
  pages = {043541},
  numpages = {32},
  year = {2017},
  month = {Feb},
  publisher = {American Physical Society},
}

@article{Berezhiani-23,
  title = {Thermalization, fragmentation, and tidal disruption: the complex galactic dynamics of dark matter superfluidity},
  author = {Berezhiani, Lasha and Cintia, Giordano and Khoury, Justin},
  journal = {Phys. Rev. D},
  volume = {107},
  issue = {12},
  pages = {123010},
  numpages = {17},
  year = {2023},
  month = {Jun},
  publisher = {American Physical Society},
}

@article{Ferreira-21,
  title = {Ultra-light dark matter},
  author = {Ferreira, Elisa G. M.},
  journal = {The Astronomy and Astrophysics Review},
  volume = {29},
  issue = {7},
  pages = {1-186},
  numpages = {},
  year = {2021},
  month = {September},
  publisher = {Springer Nature},
  issn={1432-0754}
}

@ARTICLE{Matos-24,
AUTHOR={Matos, Tonatiuh  and Ureña-López, Luis A.  and Lee, Jae-Weon }, 
TITLE={Short review of the main achievements of the scalar field, fuzzy, ultralight, wave, BEC dark matter model},      
JOURNAL={Frontiers in Astronomy and Space Sciences},        
VOLUME={11},
YEAR={2024},
ISSN={2296-987X}
}

@article{Brito-15,
title = {Black holes as particle detectors: evolution of superradiant instabilities},
author = {Brito, Richard and Cardoso, Vitor and Pani, Paolo},
journal = {Classical and Quantum Gravity},
year = {2015},
month = {jun},
publisher = {IOP Publishing},
volume = {32},
number = {13},
pages = {134001}
}

@article{Arvanitaki-17,
  title = { {Black hole mergers and the QCD axion at Advanced LIGO}},
  author = {Arvanitaki, Asimina and Baryakhtar, Masha and Dimopoulos, Savas and Dubovsky, Sergei and Lasenby, Robert},
  journal = {Phys. Rev. D},
  volume = {95},
  issue = {4},
  pages = {043001},
  numpages = {6},
  year = {2017},
  month = {Feb},
  publisher = {American Physical Society}
}

@article{Ng-21,
  title = {Constraints on Ultralight Scalar Bosons within Black Hole Spin Measurements from the {LIGO-Virgo GWTC-2}},
  author = {Ng, Ken K. Y. and Vitale, Salvatore and Hannuksela, Otto A. and Li, Tjonnie G. F.},
  journal = {Phys. Rev. Lett.},
  volume = {126},
  issue = {15},
  pages = {151102},
  numpages = {8},
  year = {2021},
  month = {Apr},
  publisher = {American Physical Society}
}

@article{LIGO-22,
  title = {All-sky search for gravitational wave emission from scalar boson clouds around spinning black holes in {LIGO O3 data}},
  author =  {},
  collaboration = {The LIGO Scientific Collaboration, the Virgo Collaboration, and the KAGRA Collaboration},
  journal = {Phys. Rev. D},
  volume = {105},
  issue = {10},
  pages = {102001},
  numpages = {28},
  year = {2022},
  month = {May},
  publisher = {American Physical Society},
}

@article{Schwarz-88,
  title = {Three-dimensional vortex dynamics in superfluid $^{4}\mathrm{He}$: Homogeneous superfluid turbulence},
  author = {Schwarz, K. W.},
  journal = {Phys. Rev. B},
  volume = {38},
  issue = {4},
  pages = {2398--2417},
  numpages = {0},
  year = {1988},
  month = {Aug},
  publisher = {American Physical Society}
}

@article{Zhao-20,
author = {Mauser, Norbert J. and Zhang, Yong and Zhao, Xiaofei},
title = {On the rotating nonlinear {Klein-Gordon} equation: nonrelativistic limit and numerical methods},
journal = {Multiscale Modeling \& Simulation},
volume = {18},
number = {2},
pages = {999-1024},
year = {2020},
}

@article{East-22,
  title = {Vortex string formation in black hole superradiance of a dark photon with the {Higgs} mechanism},
  author = {East, William E.},
  journal = {Phys. Rev. Lett.},
  volume = {129},
  issue = {14},
  pages = {141103},
  numpages = {7},
  year = {2022},
  month = {Sep},
  publisher = {American Physical Society},
}

@unpublished{Jin-25,
  author       = {Jin, Shilong and Zhao, Xiaofei and Zhang, Yong and Xiong, Chi},
  title = {},
  year = {},
  note = {in preparation}
}

@article{Davis-89,
  title = {Global strings and superfluid vortices},
  author = {Davis, R. L. and Shellard, E. P. S.},
  journal = {Phys. Rev. Lett.},
  volume = {63},
  issue = {19},
  pages = {2021--2024},
  numpages = {0},
  year = {1989},
  month = {Nov},
  publisher = {American Physical Society}, 
}

@ARTICLE{Liu-21,
  author={Liu, Daoming and Xiong, Chi and Liu, Xiaopei},
  journal={IEEE TVCG}, 
  title={Vectorizing Quantum Turbulence Vortex-Core Lines for Real-Time Visualization}, 
  year={2021},
  volume={27},
  number={9},
  pages={3794-3807}
}

@article{Vilenkin-23,
  title = {Simulating cosmic string loop captured by a rotating black hole},
  author = {Deng, Heling and Gruzinov, Andrei and Levin, Yuri and Vilenkin, Alexander},
  journal = {Phys. Rev. D},
  volume = {107},
  issue = {12},
  pages = {123016},
  numpages = {14},
  year = {2023},
  month = {Jun},
  publisher = {American Physical Society},
}

@article{Yuan-09,
    author = {Yuan, Feng and Lin, Jun and Wu, Kinwah and Ho, Luis C.},
    title = {A magnetohydrodynamical model for the formation of episodic jets},
    journal = {MNRAS},
    volume = {395},
    number = {4},
    pages = {2183-2188},
    year = {2009},
    month = {05},
    issn = {0035-8711}
}

@article{XRISM-25,
author={Collaboration XRISM },
title={Structured ionized winds shooting out from a quasar at relativistic speeds},
journal={Nature},
year={2025},
month={May},
day={01},
volume={641},
number={8065},
pages={1132-1136},
issn={1476-4687},
}

\newpage

\end{document}